\documentclass[12pt,letter]{article}
\usepackage{amsfonts,amsthm,amsmath,amssymb,graphicx,hyperref,youngtab}
\usepackage{cite}
\usepackage{mathrsfs}
\usepackage[margin=1in]{geometry}
\usepackage{graphicx,slashed,feynmf}
\usepackage{hyperref}
\usepackage{textcomp}
\usepackage{latexsym}
\usepackage{epsf}
\usepackage{cite}
\usepackage{verbatim}
 \usepackage{bbold}
 \usepackage{units}

\def\a'{\alpha'}

\def \w {\widetilde{\mathscr{W}}}
\def\mr#1{\mathrm{#1}}

\newcommand{\be}{\begin{equation}}
\newcommand{\ee}{\end{equation}}
\newcommand{\beq}{\begin{eqnarray}}
\newcommand{\eeq}{\end{eqnarray}}
\newcommand{\bl}{\noindent $\bullet$ \ }

\def\bs#1{\boldsymbol#1}

\begin{document}
\renewcommand{\thefootnote}{\fnsymbol{footnote}}

\begin{center}

\begin{flushright}

{\small WITS-CTP-133
 \\
} \normalsize
\end{flushright}
\vspace{0.9 cm}

{\LARGE \bf{Power-law Solutions from Heterotic Strings\\

\vspace{0.2cm}}}

\vspace{1.1 cm} {Tibra Ali$^{a,}$\,\footnote[4]{\tt{tibra.ali@pitp.ca}}\;\;\; S. Shajidul Haque$^{b,}$}\,\footnote[3]{{\tt{shajid.haque@wits.ac.za }}}\\

\vspace{0.9 cm}\vspace{0.2 cm}{{\it$^{a}$Perimeter Institute for Theoretical Physics \\31 Caroline Street N., Waterloo, ON N2L 2Y5, Canada} }

\vspace{0.5 cm}{{\it$^{b}$NITheP, School of Physics and Centre for Theoretical Physics\\University of the Witwatersrand, Johannesburg, WITS 2050, South Africa }} \\

\thispagestyle{empty}
\vspace{4cm}
{\bf Abstract}
\end{center}
\begin{quotation}
\noindent In this paper we search for accelerating power-law and ekpyrotic solutions in heterotic string theory with NS-NS fluxes compactified on half-flat and generalized half-flat manifolds. We restrict our searches to the STZ sector of the theory. We also considered linear order $\alpha'$ corrections for the half-flat case. The power-law solutions that we find are neither accelerating nor ekpyrotic in any of the models.
\end{quotation}

\setcounter{page}{0}
\setcounter{tocdepth}{2}
\newpage

\tableofcontents
\renewcommand{\thefootnote}{\arabic{footnote}}
\setcounter{footnote}{0}
\section{Introduction}
The classic phenomenology scenario derived from string theory is that of $\mr{E}_8\times \mr{E}_8$ heterotic strings compactified on Calabi-Yau manifolds \cite{Candelas1}, but these models suffered from a plethora of problems including supersymmetry breaking, zero cosmological constant and too many massless moduli. In recent years some of these problems have been addressed in the context of type II strings by introducing sources and turning on background fluxes on Calabi-Yau manifolds \cite{Grana}. However, in the heterotic setting there is a dearth of sources and fluxes, and an alternative to turning on fluxes is to replace the compactification Calabi-Yau manifold by SU(3) manifolds known as \emph{half-flat mirror manifolds}, which bear close resemblance to the original Calabi-Yau manifolds \cite{Gurrieri0, Gurrieri1, Gurrieri2}.

In this paper we consider some simple models, known as STZ models in the supergravity literature, that arise from the compactification of heterotic strings on half-flat and generalized half-flat manifolds that have one complex structure and one K\"ahler structure moduli.\footnote{The meaning of ``moduli" in the half-flat and generalized half-flat cases is a bit subtle, and it is explained in the appendix of this paper.} These manifolds have SU(3) structure and are the correct generalizations of Calabi-Yau manifolds which give the desired minimal supersymmetry in four dimensions. Due to the fact that they encode tunable fluxes in their geometry,  these manifolds also allow for ways of generating potentials for the moduli fields in the effective low energy theory. In particular, we compute the low-energy potentials for the STZ fields that are derived from the Gukov-Vafa-Witten type superpotentials  of \cite{Gurrieri2,deCarlos}.  These models are consistent with $\mathcal{N}=1$ supergravity.

Here we are interested in investigating whether these models yield cosmological power-law solutions, also known as scaling solutions. These are cosmological attractor solutions. For example, an exponential potential has a scale factor that satisfies power-law behaviour. Similar analyses have been done in \cite{Blaback:2013sda} for maximal ($\mathcal{N}=8$) and minimal ($\mathcal{N}=1)$ supergravity theories, based on earlier works \cite{Rosseel, Townsend}. Although our main motivation was to look for late-time accelerating solutions we also look for ekpyrotic power-law cosmologies. The reason for looking for these two classes of solutions is that the mathematical techniques for doing so are very similar. In our calculations looking for accelerating solutions we also include first order $\alpha'$ corrections to the effective action.

 To be more specific, we search for accelerating and ekpyrotic power-law cosmologies in some simple heterotic supergravity models. For these supergravity theories the dynamics of the scalar sector is dictated by a scalar potential which has the form
\be\label{eq:multiple_exponentials}
V (\vec \phi)=\sum_{a} \Lambda_a\ e^{\vec \alpha_a \cdot \vec \phi} ,
\ee
where $\Lambda_a$ and $\vec{\alpha}_a$ are real numbers\footnote{$\Lambda_a$ can in general depend on the all the scalar fields in the theory.}, and $\vec{\phi}$ is a vector consisting of scalar fields. To realize the power-law behaviour, we want to rewrite the scalar potential in the following form
\begin{equation} \label{eq:single_exponential}
V = e^{c \, \psi} \, U \, ,
\end{equation}
where $\psi$ is the running scalar field and $U$ is a function of the remaining scalar fields which need to be stabilized. The coefficient  $c$ in the exponent of the above expression is related to the power-law behaviour $P=\nicefrac{2}{c^2}$, where the scale factor goes as $a (t) \sim t^P$. There are several types of scaling solutions. The scaling solution will be accelerating when $P > 1$ (or $c^2<2$) and $U$ is stabilized at a positive value \cite{Collinucci:2004iw, Hartong:2006rt}. However, the scaling solution will be ekpyrotic\footnote{These critical values are convention dependent. Our convention is that of \cite{Rosseel} and is explained in section \ref{sec:stz-on-hf}.} when $P < \nicefrac{1}{3}$ (or $c^2>6$) and $U$ is stabilized at a negative value \cite{Khoury:2004xi}. In this paper we concentrate on these two types of scaling solutions. 
Also, we study heterotic string theory which is phenomenologically very interesting but has been much less studied in the context of cosmology.

Inflation suggests that the very early universe went through an enormous amount of acceleration. Within a fraction of a second, it grew from a subatomic scale to a macroscopic scale. Apart from conceptual issues with inflation, it would appear that for the first time some inflationary models seem to be disfavoured \cite {Ijjas:2013vea, Ijjas:2014nta} by the recent data from WMAP, ACT and Planck2013. Ekpyrotic or cyclic cosmology \cite{Steinhardt:2001st, Khoury:2001bz, Khoury:2003rt} is the most popular alternative to the theory of cosmic inflation. It proposes that our universe is going through an infinite number of cycles. Each of these cycles begins with a big bang phase followed by a slowly accelerating expansion phase, which in turn is followed by a slow contraction phase, finally ending with a big crunch. The scaling behaviour occurs during the slow contraction phase of the evolution of the ekpyrotic universe. This corresponds to a steep and negative minimum potential, since at the end of this contraction phase the universe bounces back to a positive value. This scaling phase of the ekpyrotic solutions can be realized in supergravity theories \cite{Blaback:2013sda}. In this paper, we search for this part of ekpyrotic universes for some simple heterotic supergravity theories.

On the other hand, accelerating power-law solutions are the most popular alternative to de Sitter solutions as models for late time acceleration. Most of the research on the construction of de Sitter \cite{Silverstein:2007ac, Haque:2008jz,Caviezel:2008tf, Flauger:2008ad, Danielsson:2009ff, Danielsson:2010bc, Danielsson:2011au} and power-law solutions \cite{Blaback:2013sda}  is done in the context of type II supergravity. Moreover, all known de Sitter critical points \cite{deRoo:2002jf, deRoo:2003rm, deRoo:2006ms, Gibbons:2001wy, Cvetic:2004km} develop some instability in the scalar spectrum.  One can get de Sitter solutions without instabilities in type II theories only by including either non-geometric fluxes \cite{Damian:2013dwa, Blaback:2013ht,Danielsson:2012by} or by considering non-perturbative corrections \cite{Blaback:2013qza,Danielsson:2013rza}. Consequently, in this context it is natural to study the best alternative to de Sitter which are the power-law solutions for other string theories such as the heterotic string theory.

Another motivation for looking for cosmological solutions other than de Sitter spaces is the no-go theorem of \cite{Green} excluding the existence of de Sitter vacua in compactification of heterotic strings. Note that this no-go theorem takes into account the leading order $\alpha'$ corrected heterotic string action and so is very different from the usual supergravity no-go theorem \cite{Gibbons,Maldacena} that is invoked to exclude de Sitter spaces which only works for supergravity. However, \cite{Green} only considers a time independent dilaton and is agnostic about how time-dependent model-dependent scalar fields influence the cosmology. Although we don't look for de Sitter spaces we do assume time dependence of our scalar field, and so our work is somewhat tangentially related to that of \cite{Green}.

Heterotic string theory is considered to be the most attractive string theory from the perspective of particle physics phenomenology. Yet from the cosmological perspective, it still remains largely unexplored. Part of the reason is the absence of the RR fluxes and sources, which makes it less flexible from the model building angle. Thus it is natural to consider manifolds which are generalizations of Calabi-Yau manifolds as they allow one to play with more ingredients. For example, in this paper we consider manifolds with SU(3) structure which have non-vanishing intrinsic torsion. In particular we consider half-flat mirror manifolds where the intrinsic torsion measures the deviation of these manifolds from Calabi-Yau manifolds. These torsion classes come with geometric fluxes, which play an important role in moduli stabilization. For our purposes they become additional parameters that one can then play with. We also consider leading $\alpha'$ correction for these mirror half-flat manifolds which potentially provides us with more room to play with.

Finally, we explore an even more general class of manifolds with non-integrable $\mr{SU}(3)\times \mr{SU}(3)$ structure, described in \cite{  D'Auria:2004tr}, which in \cite {deCarlos} are referred to as the generalized half-flat manifolds. These generalized half-flat manifolds do not seem to have an obvious geometric interpretation but their relevant properties can be encoded in certain simple equations that enter our calculations, as we describe in our appendix. For related work in moduli stabilization in the heterotic setting see \cite{Gray,Klaput,Lukas}.

The paper is organized as follows. In section 2 we discuss the STZ models and the K\"ahler potential and superpotential for these models for compactification on half-flat and generalized half-flat manifolds. In section 3 we search for ekpyrotic and accelerating power-law solutions for the STZ model derived from compactification on half-flat manifolds. In doing so we also briefly explain our methodology. We conclude this section by considering the $\alpha'$ correction for the half-flat case and investigate the power-law solutions. In section 4 we examine generalized half-flat manifolds and we find two ekpyrotic solutions. We do not examine the full $\a'$ corrected potential for the generalized half-flat manifolds as the full expression for the potential is extremely lengthy and unwieldy. We conclude with a summary of the paper and with some open problems. In the appendix we briefly explain mirror half-flat and generalized half-flat manifolds. We also describe the superpotential and the K\"ahler potential used in our calculations.

\section{STZ Model: K\"ahler Potential and Superpotential}
Although our ultimate goal is to study the heterotic string cosmology that comes from compactification on half-flat and generalized half-flat manifolds, we restrict our attention in this paper to the minimal truncation of the four-dimensional effective theories that arise from these types of compactifications. These models are known as STZ models in the supergravity literature and they arise from restricting to one modulus each from both the complex structure and K\"ahler structure moduli spaces -- the corresponding moduli are Z and T, respectively. On the other hand the complex scalar field S is the axion-dilaton field which is universal in all of these models as they correspond to the four dimensional $H$ field (which becomes a scalar upon dualization) and the four dimensional component of the dilaton. In this section we collect our master formulae, partly to establish our conventions.

\subsection{The K\"ahler Potential} \label{kahler}
The expression for the K\"ahler potential and the superpotential, including the $\alpha'$ corrections, are taken from \cite{Gurrieri2}. The details for the K\"ahler potential are given in the appendix \ref{sec:kahler}, and here we quote the results. The K\"ahler potential for the STZ model at the zeroth order in $\alpha'$ is given by
\begin{equation}
K = K_{\mathrm{cs}} + K_{\mathrm{K}} + K_{S}\ ,
\end{equation}
with 
\beq
K_{\mathrm{cs}} = - 3 \ln \{ i (\bar{Z}-Z)\}, \
K_S =- \ln \{i (\bar{S}-S)\}, \ \text{and} \ 
K_{\mathrm{K}} = - 3 \ln \{i(T-\bar{T})\}.
\eeq
Where $K_{\mr{cs}}, K_{S}$ and $K_{\mr{K}}$ are the K\"ahler potentials arising from the complex structure modulus, the axion-dilaton, and the K\"ahler structure modulus, respectively. On the other hand, the first order in $\a'$ correction to the K\"ahler potential is given by,
\begin{align}
K_{\alpha'}&= -3\alpha'\left(\frac{C\bar{C}}{(Z-\bar{Z})^2} + \frac{4D\bar{D}}{(T-\bar{T})^2}+ \frac{6 (CD+ \bar{C}\bar{D})}{(T-\bar{T})(Z-\bar{Z})}\right),
\end{align}
where $C$ and $D$ denote matter fields in the K\"ahler and the complex structure sectors, respectively, which belong to the $\boldsymbol{27}$ (and $\overline{\boldsymbol{27}}$) representation of $\mr{E}_6$. The $\mr{E}_6$ GUT symmetry arises as a generalization \cite{Gurrieri2, Ali} of the standard embedding of the usual Calabi-Yau scenario \cite{Candelas1}. The $\mr{E}_6$ indices have been largely suppressed in this paper. Also, throughout this paper we shall assume that these fields take on {\sc vev}s that are at the GUT symmetry breaking scale or some other suitable high energy scale. The precise details of GUT symmetry breaking is beyond the scope of this paper. 

The coefficients to the kinetic terms at the leading order that one gets from the zeroth order K\"ahler potential are given by the a K\"ahler metric which has the following components,
\be
K_{S\bar{S}} =  -\frac{1}{(S-\bar{S})^2}, \ K_{T\bar{T}}  = -\frac{3}{(T-\bar{T})^2}, \
K_{Z\bar{Z}}  = -\frac{3}{(Z-\bar{Z})^2},
\ee
with all other components vanishing. However, there are corrections to some of these kinetic terms as well as the off-diagonal terms (which are zero at the zeroth order) at the $\alpha'$ order. But we ignore these corrections when looking for accelerating scaling solutions for the following reason. When looking at accelerating solutions we imagine that we are in a slow role regime in which the zeroth order kinetic terms can be ignored. The $\a'$ correction to the kinetic terms are further suppressed by a factor of $\a'$. These two factors combine to justify ignoring the $\a'$ corrections to the K\"ahler potential. 

But in the case of ekpyrosis we cannot assume that the kinetic term for the ekpyrotic scalar is small, as it has to be of the same order of magnitude as the potential term. Thus, if we include $\a'$ corrections to the potential term we also need to include the $\a'$ corrections to the kinetic terms, and then redefine the scalars such that the new scalars are canonically normalized. Here we run into a difficulty as we find that even for the simplest STZ model, the $\a'$ corrections to T and Z fields are non-linear and so expressing the old fields in terms of the new variables becomes extremely messy without making additional assumptions regarding the values of the fields. Thus, in this paper we restrict ourselves to looking at \emph{both} the kinetic and potential terms only the zeroth order in $\a'$ when looking for ekpyrotic solutions. For completeness, the full metric to first order in $\a'$  of the K\"ahler manifold is given in appendix \ref{sec:kahler}.
\subsection{The Superpotential}
In this subsection we present the superpotential computed in \cite{Gurrieri2} and then specialize to the case where we keep only one K\"ahler modulus and one complex structure modulus as is appropriate to the STZ models.

According to \cite{Gurrieri2} the Gukov-Vafa-Witten superpotential arising from compactification on generalized half-flat manifolds is given by
\begin{align}
W= \tilde{\epsilon}_A Z^A - \tilde{\mu}^A \mathscr{G}_A, \label{eq:zero-order-superpotential}
\end{align}
with
\beq
\tilde{\epsilon}_A &=& \epsilon_A - T^i p_{Ai} \cr
\tilde{\mu}^A &=& \mu^A - T^i q^A_i.
\eeq
Let us describe the different components of these formulae. The $Z^A$ are the complex structure moduli given in projective coordinates on the moduli space. This means that one of the $Z^A$ is redundant. Consequently, in some of the following formulae we set $Z^0=1$ and denote the rest of the complex structure moduli by $Z^a$.  The $\mathscr{G}_A =\frac{\partial \mathscr{G}}{\partial Z^A}$ are the first derivative of the pre-potential $\mathscr{G}$ whose explicit form is given in the appendix \ref{sec:superpotential}.

As explained in \cite{deCarlos, Gurrieri2}, the fact that certain forms fail to close on half-flat and generalized half-flat manifolds as compared to Calabi-Yau manifolds allows us to turn on additional component of $H$ fluxes. $\epsilon^A$ and $\mu_A$ in the formulae above, defined more precisely in appendix \ref{sec:generalized half-flat}, are flux parameters that come from turning on these $H$ fluxes. On the other hand $p_{Ai}$ and $q^A_i$ are torsion parameters associated with half-flat and generalized half-flat structures of the internal manifold. The $A$ index in these parameters is associated with the complex structure moduli whereas the $i$ index is associated with the K\"ahler structure.

When both $q^A_i$ and $p_{Ai}$ are non-zero, the internal manifold is a generalized half-flat manifold. When $q^{A}_i=0$ and $p_{ai}=0$ it reduces to a half-flat manifold. In other words, the half-flat case corresponds to only the $p_{0i}\ne 0$. See appendices \ref{sec:half-flat} and \ref{sec:generalized half-flat} for more details. When all of these parameters vanish we recover the Calabi-Yau case.

Putting in the explicit expressions for $\mathscr{G}_A$ and setting $Z^0=1$, we arrive at:
\begin{align}
W= (\epsilon_0 - p_{0i}T^i)+ (\epsilon_a - p_{ai} T^i) Z^a +  \frac{1}{2} \tilde{d}_{abc} (\mu^a - q^a_i T^i) Z^b Z^c - \frac{1}{6} \tilde{d}_{abc}(\mu^0 - q^0_i T^i) Z^a Z^b Z^c ,
\end{align}
where $\tilde{d}_{abc}$ are related to the Yukawa couplings of the original Calabi-Yau manifold and has the interpretation of being the intersection numbers of the mirror dual of the Calabi-Yau manifold in the large complex structure limit. However, in the case of STZ model $\tilde{d}_{abc}$ has only one component and its value is simply 1 for the regime we are in. In this limit the superpotential becomes:
\begin{align}
W= (\epsilon_0 - p_{01}T)+ (\epsilon_1 - p_{11} T) Z +  \frac{1}{2} (\mu^1 - q^1_1 T) Z^2 - \frac{1}{6} (\mu^0 - q^0_1 T) Z^3.
\end{align}\\
The first order in $\a'$ correction to the superpotential has been computed in \cite{Gurrieri2}. For the STZ model arising from the compactification on a generalized half-flat manifold it is given by
\begin{align}
W_{\alpha'} &= 2 \alpha' \left(p_{11} -\frac{1}{2} q^0_1 Z^2 + q^1_1 Z\right) C_P D^P  - \frac{\alpha'}{3} \left\{j_{\bar{P}\bar{R}\bar{S}} C^{\bar{P}}C^{\bar{R}}C^{\bar{S}} + j_{PRS} D^P D^R D^S\right\},
\end{align}
where $j_{PRS}$ ($j_{\bar{P}\bar{R}\bar{S}}$) is the singlet piece in the $\boldsymbol{27}\times\boldsymbol{27}\times\boldsymbol{27}$ ($\boldsymbol{\overline{27}}\times\boldsymbol{\overline{27}}\times\boldsymbol{\overline{27}}$) of $\mr{E}_6$.\\

Given the K\"ahler potential and the superpotential, the scalar potential (see appendix (\ref{sugrapot}) for our convention in which we set the Planck scale $\kappa^2=1$) is given by the formula:
\be\label{potential}
V=e^{K} \left ( \sum_{\Phi} K^{I\bar J} D_{I} W D_{\bar J} \bar W -3 |W| ^2 \right) .
\ee
Here $\Phi^I$ is a generic complex scalar field ($\bar{\Phi}^{\bar{I}}$, its complex conjugate), $K^{I \bar J}$ is the inverse K\"ahler metric to $K_{I\bar J}=\frac{\partial^2 K}{ \partial \Phi^I \partial \bar{\Phi}^{J}  }$ and $D_{I} W=\frac{\partial W   }{\partial \Phi^I  }+\frac{\partial K}{\partial \Phi^I  } W$. The general expression for $V$ for generalized half-flat manifolds with $\alpha'$ correction can be easily computed. 
However, the expression is long, unwieldy and not very instructive and so we  do not reproduce it here. As usual, the kinetic terms of the complex scalar fields are given by 
\be
\mathcal{L}_{kin} =K_{I \bar J} \partial_\mu {\Phi^I} \partial^\mu {\bar \Phi}^{\bar{J}},
\ee
keeping in mind that for the kinetic terms we ignore the $\alpha'$ corrections as discussed above.
\section{STZ on Half-flat Manifold}\label{sec:stz-on-hf}
In this section we examine the scalar potential for the STZ model that arises from compactification on half-flat manifolds. For a brief review of half-flat manifolds see appendix \ref{sec:half-flat}. For now we only consider the zeroth order (in $\alpha'$) potential. The $\alpha'$ corrected potential is examined in the next subsection. The superpotential in this case is given by
\beq
W&=& - e T +\epsilon Z +\frac{\mu}{2} Z^2.
\eeq
Here $\epsilon$ and $\mu$ are flux parameters, while $e$ is a torsion parameter, that have been relabelled. We define the complex scalars in terms of real scalars as follows:
\be
S= s -i e^{\sqrt{2} \sigma} , \
T=  t -i e^{\sqrt{\frac{2}{3}}\tau} , \
Z =-z - i e^{\sqrt{\frac{2}{3}}\zeta}.
\ee
We, then, get the following scalar potential:
\beq \label{pot}
V_{scalar}&=&\left (\frac{e^2}{96} \right )\ e^{-\sqrt{6} \zeta - \sqrt{2} \sigma - \sqrt{\frac{2}{3}} \tau} +\left(\frac{\mu^2}{384}\right) e^{\sqrt{\frac{2}{3}} \zeta - \sqrt{2} \sigma - \sqrt{6} \tau} +\left\{\frac{(\epsilon - z \mu)^2}{96}\right\} e^{-\sqrt{\frac{2}{3}} \zeta - \sqrt{2} \sigma - \sqrt{6} \tau} \cr
&+& \left\{\frac{\left (2 e t + z (2 \epsilon - z \mu)\right)^2}{128} \right \} e^{-\sqrt{6} \zeta- \sqrt{2} \sigma - \sqrt{6} \tau} .
\eeq
The kinetic part of the Lagrangian becomes:
\be
\mathcal{L}_{kin}=\frac{1}{2} \left\{ (\partial \zeta)^2 + (\partial \sigma)^2+ (\partial \tau)^2 + \frac{3}{2} e^{-2 \sqrt{\frac{2}{3}}\zeta} (\partial z)^2  + \frac{1}{2} e^{-2 \sqrt{2} \sigma} (\partial s)^2 + \frac{3}{2} e^{-2 \sqrt{\frac{2}{3}}\tau} (\partial t)^2  \right\}.
\ee
The $\{\zeta, \sigma, \tau\}$ submanifold is flat and the fields are canonically normalized. \\

Since the critical values of the parameter $P$ are dependent on the choice of units, we briefly explain here our convention and review the derivation of the critical values. Let's consider the action for gravity minimally coupled to a single canonically normalized scalar field:
\begin{align}
\mathcal{S}=   \int dx^4 \sqrt{-g} \left( \frac{\mathcal{R} }{2 \kappa^2}- \frac{1}{2} \partial_\mu \varphi \partial^\mu \varphi - V(\varphi)\right),
\end{align}
where $\kappa^2=8 \pi G$, with $G$ Newton's constant.  
As explained before, we choose  $\kappa^2=1$ in this paper.  

We set $\varphi=\varphi(t)$ where $t$ is the cosmological time in the flat FRW metric, and choose the simple potential
\begin{align}
V=\Lambda e^{c\varphi},
\end{align}
with $\Lambda$ constant. Now, if we introduce the variables \cite{Hartong:2006rt},
\be
\label{eq:auto1} x =\frac{\dot{\varphi}}{\sqrt{6}H}, \
y =\frac{\Lambda e^{c\varphi}}{3H^2},
\ee
where $H=\nicefrac{\dot{a}}{a}$ is the Hubble parameter, the equations of motion can then be put in the form of an autonomous system of equations,
\begin{align}
x'&=-3 xy - \sqrt{\frac{3}{2}} \ c y\\
y'&=y(\sqrt{6}cx+6x^2)
\end{align}
plus a constraint equation,
\begin{align}
x^2+y=1.
\end{align}
The prime implies derivative with respect to $\ln{a(t)}$. The constraint is preserved by the autonomous system, as so as long as we start with the correct boundary condition the two autonomous equations (\ref{eq:auto1}) yield critical values of $c$ whenever the right hand side of these equations go to zero. For this simple case there are three critical points, one of them being $(x=-\frac{c}{\sqrt{6}},y= 1-\frac{c^2} { 6})$, which is a true solution to Einstein's equations. The solution, in terms of the original variables, is then
\begin{align}
\varphi= -\frac{2}{c} \ln t+ \frac{1}{c} \ln \left(\frac{  12-2 c^2} { c^4 \Lambda } \right ), \ a(t) \sim t ^{\nicefrac{ 2 } { c^2 }}.
\end{align}
When $\Lambda>0$, a solution only exists when $c^2<6$, and it is accelerating when $c^2<2$ or $P>1$. On the other hand, when $\Lambda<0$, a solutions only exists if $c^2>6$ or $P < \nicefrac{1}{3}$, which corresponds to the slowly contracting phase of an ekpyrotic universe. These critical values of $P$ are completely fixed for the canonically normalized scalar once our choice of units for the Planck mass is chosen. 
\subsection{Power-law Solutions}
Before we analyze the potential for possible scaling solutions we will briefly discuss the methodology. The details of this can be found in \cite{Collinucci:2004iw,Hartong:2006rt,Blaback:2013sda}. 

The potential  (\ref{pot}) can be expressed as,
\be
V (\phi)=\sum_{a=1}^4 \Lambda_a e^{\sum_{i=1}^3 \alpha_{ai} \phi_i},
\ee
where $\vec{\phi}=\{ \zeta, \sigma, \tau\}$ is a vector consisting of the scalars fields and $\vec{\alpha}_a$ are the vectors of coefficients of the scalar fields in the argument of the exponential that corresponds to the of the $a$-th term in the potential. $\alpha_{ai}$ is the $i$-th component of the vector $\vec{\alpha}_a$. In general, for potentials of this form we denote by $M$ the number of exponential terms and by $R$ the number of scalar fields that appear in the exponential. For the potential (\ref{pot}) we have $M=4$ and $R=3$. 

The motivation for writing the potential in this form is that it helps us to rewrite the potential in the form (\ref{eq:single_exponential}) which allows us to identify the $P$ dependence of the scale factor on cosmological time, \emph{i.e.}, $a\sim t^P$. However, since $R<M$ the $\vec{\alpha}_a$ vectors entering (\ref{pot}) are not all linearly independent and a scaling solution is only possible if the set of $\vec{\alpha}_a$ vectors are mutually \emph{affine}, which we define below.

Since $R<M$ we can always choose $R$ of the $\vec{\alpha}_a$ to be linearly independent and express the rest as a linear combination of this set:
\begin{align}
\vec{\alpha}_b = \sum_{a=1}^{R} c_{ba} \vec{\alpha}_a.
\end{align}
A set of $\vec{\alpha}_a$ vectors are called \emph{affine} if the coefficients $c_{ba}$ above can be chosen such that
\begin{align}
\sum_a c_{ba}= 1  \ \ \forall \ \ b=R+1, \dots, M.
\end{align}
Once we have found out the largest set of mutually affine $\vec{\alpha}_a$ vectors we can then calculate what the $P$ values are associated with that set. For the $R<M$ case, one needs to consider $\alpha_{ai}$ as an $M\times R$ matrix and then define an $R\times R$ matrix as follows
\be
B_{ij}=\sum_{a=1}^M\alpha_{ai} \alpha_{aj}, 
\ee
where $a=1,...,M $ and $i=1, ..., R$. Then we get the $P$ value from,
\be
P=2 \sum_i \left(\sum_{ja} |B^{-1}|_{ij} \alpha_{aj} \right).
\ee 
For an accelerating scaling solution we need $P>1$ and the potential needs to be stabilized at a positive minimum, and for an ekpyrotic solution we need $P<\nicefrac{1}{3}$ with the potential now minimized at a negative value. \\\\
\bl \underline{{\bf Analysis:}}\\\\
As explained in the previous section first we will collect the $\vec \alpha$-vectors. From the scalar potential (\ref{pot}) we get the following $\vec \alpha$-vectors:
\beq \vec \alpha_1&=&\{-\sqrt{6},-\sqrt{2},-\sqrt{\nicefrac{2}{3}}\},\ 
\vec  \alpha_2=\{\sqrt{\nicefrac{2}{3}},-\sqrt{2},-\sqrt{6}\}, \cr
\vec \alpha_3&=&\{-\sqrt{\nicefrac{2}{3}},-\sqrt{2},-\sqrt{6}\}, \
\vec \alpha_4=\{-\sqrt{6},-\sqrt{2},-\sqrt{6}\}.\eeq
As outlined above, 
we need to find the $P$ values for the largest common set of $\vec{\alpha}$-vectors that are mutually affine. From these $\vec \alpha$-vectors the largest $P$ we get is $1$. Thus the scalar potential will not give us any accelerating scaling solution, since there is no combination of vectors that can give us $P>1$.

However, there might still be an ekpyrotic solution. After turning off some of the flux parameters, it may be possible to decrease the value of $P$ from $1$, and stabilize the rest of the potential at a negative value. If so, we would achieve an ekpyrotic solution. Unfortunately, for the simple STZ on mirror half-flat manifold we did not find a solution that can be stabilized at a negative minimum.
\subsection{STZ with $\alpha'$ Correction on Half-flat Manifold}
When we add linear order $\alpha'$ corrections to the K\"ahler potential and the superpotential, there is the possibility that the results given above may change qualitatively. As argued in section (\ref{kahler}), we ignore the corrections to the kinetic terms as we are looking for solutions with small kinetic energy. Using the $\alpha'$ corrections to the superpotential and the K\"ahler potential derived in \cite{Gurrieri2} and quoted above, it is a straightforward exercise to compute the $\alpha'$ correction to the scalar potential for half-flat manifolds. It is given by
\begin{align}
V_{\alpha'} &= \alpha' \left\{ - \frac{e}{96} \mathrm{Im}(J) e^{-\sqrt{6} \zeta - \sqrt{2} \sigma - 2 \sqrt{\frac{2}{3}}\tau} - \frac{z\mu}{96}\mathrm{Im}(J) e^{-2\sqrt{\frac{2}{3}} \nonumber \zeta-\sqrt{2}\sigma - \sqrt{6} \tau}\right.\\ 
 &\left. +\left[ \frac{et}{48} \mathrm{Re}(J)- \frac{\mu z^2}{96} \mathrm{Re}(J)\right] e^{-\sqrt{6}\zeta-\sqrt{2}\sigma - \sqrt{6}\tau} \right\} ,
\end{align}
where $J\equiv \left\{j_{\bar{P}\bar{R}\bar{S}} C^{\bar{P}}C^{\bar{R}}C^{\bar{S}} + j_{PRS} D^P D^R D^S\right\}$, and we have only included terms that are linear in the flux and torsion parameters in $V_{\alpha'}$. 
\subsubsection* {Accelerating Power-law Solutions}
The $\alpha'$ correction adds three new terms with two new $\vec \alpha$ vectors to the potential but still we get the same scale factor as before:
$P=1.$ So we get only generic scaling solutions.
\section{STZ on Generalized Half-flat}
Generalized half-flat manifolds were first proposed in \cite{D'Auria:2004tr}. These manifolds generalize half-flat manifolds by incorporating magnetic flux as well as electric flux in their curvature. For our purposes they are defined by the failure of certain forms to be closed, as in the half-flat case. More detailed discussion about these spaces can be found either in the appendix or the references \cite{D'Auria:2004tr,Gurrieri2}.

The K\"ahler form for the generalized half-flat compactification is the same. As mentioned before, for the STZ case we find that the superpotential is given by:
\begin{align}
W= (\xi - e T)+ (\epsilon - p T) Z +  \frac{1}{2} (\mu - q T) Z^2 - \frac{1}{6} (\rho - r T) Z^3,
\end{align}
where the flux parameters $\xi, \epsilon, \mu, \rho$ and the torsion parameters $e, p, q, r$ are subject to two sets of constraints. However, the first set of these constraints (\ref{constraint1}) is trivially satisfied for the STZ model, the second set (\ref{constraint2}) reduces to
\be \label{cons}
-\xi \ r - \epsilon \ q + \mu \ p + \rho \ e = 0 .
\ee
 The full potential for the generalized half-flat manifold is significantly more complicated than the mirror half-flat case and we don't reproduce it here. By using (\ref{potential}) we can find this scalar potential and it has nine terms with the above mentioned eight flux and torsion parameters. 
\subsection{Power-law Solutions}
As pointed out in mirror half-flat case, to illustrate the scaling behaviour we will write down the nine $\vec \alpha$-vectors. They are as follows:\\\\
\bl \underline{{\bf {$\vec\alpha$}-Vectors:}} \footnote{ $\vec \alpha_3,\vec \alpha_5,\vec \alpha_6,\vec \alpha_7$ are the $\vec \alpha$-vectors we have found for the mirror half-flat case.}
\beq  
\vec \alpha_1&=&\{0,-\sqrt{2},-2\sqrt{\nicefrac{2}{3}}\}, \ \ \ \ 
\vec  \alpha_2=\{-\sqrt{\nicefrac{2}{3}},-\sqrt{2},-\sqrt{\nicefrac{2}{3}}\}, \
\vec \alpha_3=\{-\sqrt{\nicefrac{2}{3}},-\sqrt{2},-\sqrt{6}\}, \cr
\vec \alpha_4&=&\{\sqrt{\nicefrac{2}{3}},-\sqrt{2},-\sqrt{\nicefrac{2}{3}}\}, \
\vec \alpha_5=\{\sqrt{\nicefrac{2}{3}},-\sqrt{2},-\sqrt{6}\}, \ \ \ \ \ 
\vec  \alpha_6=\{-\sqrt{6},-\sqrt{2},-\sqrt{\nicefrac{2}{3}}\}, \cr
\vec \alpha_7&=&\{-\sqrt{6},-\sqrt{2},-\sqrt{6}\}, \ \ \
\vec \alpha_8=\{\sqrt{6},-\sqrt{2},-\sqrt{\nicefrac{2}{3}}\}, \ \ \ \ \
\vec \alpha_9=\{\sqrt{6},-\sqrt{2},-\sqrt{6}\}. 
\eeq
These vectors are affinely connected. The largest $P$ value once again is $1$, which rules out the possibility of getting an accelerating scaling solution. However, we might get ekpyrotic solutions by turning off flux/torsion parameters. For ekpyrotic solutions we need the scale factor $P<\nicefrac{1}{3}$ and the potential is stabilized at a negative minimum. Since for the full potential $P=1$ we need to turn off fluxes to lower the $P$ value. However, we did not find any solutions with $P<\nicefrac{1}{3}$.

As explained before, we did not look at the full $\a'$ corrected potential for generalized half-flat manifolds in this paper due to the fact that it is difficult, even for the simplest STZ models, to diagonalize the kinetic terms when $\a'$ corrections are taken into account. 
 
\section{Discussion}
In this paper, we have investigated cosmological power-law solutions from the $\mr{E}_8\times \mr{E}_8$ heterotic superstring theory. Our goal was to search for power law solutions -- both accelerating and ekpyrotic -- as alternatives to de Sitter vacua and inflation, respectively. With this aim we explored the STZ models which are the simplest models that can arise from heterotic strings with NS-NS fluxes compactified on mirror half-flat and generalized half-flat manifolds with only one complex structure and one K\"ahler structure moduli. 

For the STZ model on mirror half-flat manifolds we found that there are four exponential terms in the full scalar potential. We searched for accelerating solutions but we didn't find any. Due to the fact that the half-flat mirror manifold has nonzero intrinsic torsion, the theory has geometric fluxes. These fluxes can be turned off and we tried to reduce the $P$-value by turning off fluxes and searched for ekpyrotic solutions. We did not find any such solutions either. We also calculated the $\alpha'$ correction to the potential which adds three extra terms in the scalar potential. However, including the $\alpha'$ correction to the potential did not change the situation.

We then considered a much more generic scenario which arises from compactification on generalized half-flat manifolds. We restricted ourselves to manifolds which give rise to STZ models and hence we still have the minimal six real scalar fields. But importantly for the generalized half-flat manifolds, we have more fluxes and torsion parameters to play with and consequently the full potential we find was much more complicated. To be precise, the full scalar potential has nine exponential terms with eight flux and torsion parameters at zeroth order in $\a'$. Using the same method as before we searched for both accelerating and ekpyrotic solutions but did not find any.
Due to the complicated nature of the kinetic terms when $\a'$ corrections are included we have not explored how these solutions are changed once these corrections are taken into account.

Before we close this paper let us make a few brief comments about possible extensions. We haven't explored the full $\a'$ corrected potential for the generalized half-flat manifold due to its  complicated nature. The next step would be to explore the full potential to see if one can obtain accelerating solutions or ekpyrotic solutions. Another obvious extension of this work would be to consider more realistic half-flat and generalized half-flat manifolds which have more than one moduli in either the complex structure or the K\"ahler structure sectors or both. The structure of the potential would be much more complicated and it might give rise to an interesting dynamics that may include different cosmological phases. Another ingredient that one may consider are instanton corrections to the potential coming from world-sheet instantons. Next, since we want to make connection between cosmology and particle physics, as should be natural in the heterotic setting, one should consider the dynamics of the $\mr{E}_6$ matter field as well. 

\section*{Acknowledgement} 
We would like to thank J. Bl{\aa}b\"ack and S. Nampuri for reading the draft and giving us some useful feedback. We also thank Ilies Messamah for discussions. The work of SSH is supported by the South African Research Chairs Initiative of the Department of Science and Technology and National Research Foundation. SSH also would like to thank Perimeter Institute for their hospitality where part of the work was done. Research at Perimeter Institute is supported by the Government of Canada through Industry Canada and by the Province of Ontario through the Ministry of Economic Development and Innovation.
\appendix

\section{Appendix} 
\subsection{The Supergravity Potential} \label{sugrapot}
The scalar sector of 4d $\mathcal{N}=1$ supergravity Lagrangian is given by
\be
\mathcal{L}=- \sum_{\Phi} K_{I\bar{J}} \partial_\mu \Phi^i \partial^\mu \Phi^{\bar{j}}- V_F
\ee
where $K_{I\bar{J}}=\frac{\partial^2 K}{\partial \Phi^I \partial \Phi^{\bar{J}}}$ is the K\"ahler metric on the moduli space and it is derived from the K\"ahler potential $K(\Phi^I, \Phi^{\bar{J}})$. The scalar potential $V_F$ is the $F$-term potential and it is derived from $K$ and the holomorophic superpotential $W(\Phi^I)$ from,
\be
V_F = e^{\kappa^2 K}\left\{ K^{I\bar{J}} D_I W D_{\bar{J}} \bar{W} - 3 \kappa^2 |W|^2\right\},
\ee
where $K^{I\bar{J}}$ is the inverse metric on the K\"ahler moduli space and the covariant derivative is defined by $D_I W\equiv \frac{\partial W}{\partial \Phi^I} + \kappa^2 \frac{\partial K}{\partial \Phi^I} W$. In this paper we follow the convention $\kappa^2=1$.
\subsection{Notes on half-flat Manifolds}\label{sec:half-flat}
In order to obtain a four-dimension theory with $\mathcal{N}=1$ supersymmetry, the most general manifold on which one can compactify heterotic string theory is a six dimensional manifold with SU(3) structure. Such manifolds are known as SU(3) manifolds. Half-flat manifolds happen to be a subclass of SU(3) manifolds. 

Recall that a generic six dimensional Riemannian manifold will have SO(6) as its structure group. This means that on overlaps of coordinate patches the tangent bundle is sewn together by transition functions which take their values in SO(6). SU(4) is the double cover of SO(6) and, when the structure group is reduced from the generic SU(4) down to SU(3), this implies restrictions on the holonomy group and hence the curvature tensor which generates that holonomy group.

However, this curvature tensor need not be the Riemann curvature tensor which generates the holonomy group of the torsion-free Levi-Civita connection which we denote by $\nabla$. For a generic SU(3) manifold the curvature tensor that is associated with the holonomy group will be the one associated with a \emph{torsionful} connection $\nabla'$. The deviation of this connection from the Levi-Civita connection can be shown to be parametrized by a quantity called ``the intrinsic torsion", defined by $T=\nabla'-\nabla$. Since the manifold has SU(3) as its structure group, it is natural to classify the intrinsic torsion in terms of representations of SU(3). These are called SU(3) modules and it can be shown \cite{Chiossi:2002tw} that under SU(3) the intrinsic torsion breaks up into five modules given by

\begin{equation}
\begin{array}{ccccccccccccccccccc}
(\bs{1} &\oplus & \bs{1}) &\oplus & (\bs{8} &\oplus & \bs{8}) &\oplus &(\bs{6} &\oplus &\bar{\bs{6}}) &\oplus &(\bs{3} & \oplus & \bar{\bs{3}}) &\oplus & (\bs{3'} & \oplus & \bar{\bs{3}}'),\\
&W_1 & &\oplus & &W_2 & &\oplus & &W_3 & &\oplus & &W_4 & &\oplus & &W_5 \nonumber 
\end{array}
\end{equation}
where $\bs{3}$ and $\bs{3}'$ represent two different SU(3) modules. These $W_i$s are known as  torsion classes and they encode the deviation of the Levi-Civita connection $\nabla$ from SU(3) holonomy. In string theory language $W_i$s encode the deviation of from Calabi-Yau manifold while retaining supersymmetry. A manifold is called \emph{half-flat} if the following conditions on intrinsic torsion hold:
\begin{equation}
W^-_1 = W^-_2 =W_4 =W_5 =0.
\end{equation}
Here $W^-_{i}$ denote the imaginary part of the torsion class. A Calabi-Yau manifold is an SU(3) manifold where all the torsion classes vanish.

Half-flat manifolds arise naturally in string theory \cite{Gurrieri0} in the following way. Type IIA(B) theory on a Calabi-Yau $\mathscr{M}$ with non-zero ``electric" components of $H$ flux is mirror dual to type IIB(A) theory on half-flat manifold $\widetilde{\mathscr{W}}$ without any NS-NS or R-R fluxes. $\widetilde{\mathscr{W}}$ can be considered to be a deformation of the ``canonical" mirror $\mathscr{W}$, \emph{i.e.},\ the Calabi-Yau mirror of $\mathscr{M}$ in the absence of $H$ fluxes. Half-flat manifolds that arise in this way are called \emph{half-flat mirror manifolds} and these are the class of manifolds that interests us from a phenomenological point of view.

For our purposes half-flat mirror manifold $\widetilde{\mathscr{W}}$ can be defined in terms of the failure of closure of $\omega_i$ and, $\alpha_A$ and $\beta^A$ which form the basis for the second and third cohomology groups of $\mathscr{W}$. Recall that in the Calabi-Yau case these forms were all harmonic forms and hence closed. In particular they satisfy the relationships:
\begin{equation}
\begin{split}
d\omega_i &= d\tilde{\omega}^i = d\alpha_A= d\beta^A =0,
\end{split}
\end{equation}
where $\omega_i$ form the basis of $H^{1.1}(\mathscr{W})$ and $\tilde{\omega}^i$ form the basis of the Hodge-dual cohomology group $H^{2,2}(\mathscr{W})$ with $i=1,\dots, h^{1,1}$ where $h^{1,1}$ is the dimension of the cohomology group $H^{1,1}(\mathscr{W})$. Similarly the set $\{\alpha_A, \beta^B\}$ with $A,B=0,1,\dotsm h^{1.2}$ denote the symplectic basis for $H^{1,2}(\mathscr{W}) \oplus H^{2,1}(\mathscr{W})$, where $h^{1,2}$ is the dimension of $H^{1,2}(\mathscr{W})$. Note that $\{\alpha_A, \beta^B\}$ is a projective basis as the range that $A$ and $B$ take are from 0 to $h^{2,1}$. The corresponding moduli fields are denoted by $Z^A$. In our calculations we shall set $Z^0=1$ after completing the calculations. We adopt the convention that lower-case Latin alphabet $a,b$ range from 1 to $h^{1,2}$, i.e. excludes the ``unphysical" component $Z^0$. For details regarding the formalism of Calabi-Yau manifolds see \cite{Candelas2}.

The basis introduced above satisfy the very important relationships:
\begin{equation}
\int \alpha_A \wedge \beta^B = \delta_A^B, \ 
\int \alpha_A \wedge \alpha_B = 0, \ 
\int \beta^A \wedge \beta^B = 0
\end{equation}
and 
\begin{equation}
\int\omega_i \wedge \tilde{\omega}^j = \delta_i^j .
\end{equation}
In the half-flat deformation $\widetilde{\mathscr{W}}$ these elements fail to close in the following way:
\begin{equation}
d\alpha_0 = e_i \tilde{\omega}^i \;\;\;\; d\alpha_a =d\beta^A=0\;\;\;\; d\omega_i = e_i \beta^0\;\;\;\; d\tilde{\omega}^i=0 , \label{eq:half-flat}
\end{equation}
where $e_i$ are the torsion parameters.

Recall the definition of a Calabi-Yau manifold is that it is a K\"ahler manifold with vanishing first Chern class. These conditions imply that the K\"ahler form $J$ is closed and that there exists a closed holomorphic three form $\Omega$:
\begin{equation}
dJ = d\Omega =0.
\end{equation}
In terms of the bases introduced above $J$ and $\Omega$ can be expressed as
\begin{equation}
\begin{split}
J & = t^i \omega_i\\
\Omega &= Z^A \alpha_A - \mathscr{G}_A(Z) \beta^A ,
\end{split}
\end{equation} 
where $t^i$ is the K\"ahler moduli field which gets complexified from a contribution from the NS-NS two form potential $B=\tau^i \omega_i$ and $Z^A$ are the complex structure moduli. $\mathscr{G}_A(Z) = \frac{\partial\mathscr{G}(Z)}{\partial Z^A}$ is the derivative of the prepotential $\mathscr{G}(Z)$ on the moduli space of $\mathscr{W}$.

From the rules (\ref{eq:half-flat}) that define half-flat manifolds we then get for $\w$ the following relationships
\begin{equation}\begin{split}
dJ & = e_i t^i \beta^0\\
d\Omega &= e_i \tilde{\omega}^i .
\end{split}
\end{equation}
In the case of Calabi-Yau compactification the fields $v^i$ and $Z^a$ are interpreted as moduli fields. As the half-flat manifolds can be thought of as deformations of an underlying Calabi-Yau manifold, these fields continue to have the same interpretation. However, due to the fact that the internal manifold is no longer Ricci-flat gives rise to potential in the low energy effective theory that contain these fields. 
\subsection{Notes on Generalized Half-flat Manifolds} \label{sec:generalized half-flat}
In this paper we also look at effective potentials arising from compactifications of heterotic strings on generalized half-flat manifolds. These manifolds are examples of manifolds with generalized complex structures introduced by Hitchin \cite{Hitchin:2004ut}. Generically these spaces have (non-integrable) $\mr{SU}(3)\times \mr{SU}(3)$ structure \cite{Grana:2005ny}. But here we concentrate on generalized half-flat manifolds which can be thought of as manifolds on which one has magnetic fluxes turned on in addition to the usual electric fluxes. It is not clear what the geometrical interpretation of these manifolds is but for our purposes we take their working definition to be given in a way analogous to half-flat manifolds, namely by the failure of certain forms to close \cite{Gurrieri2}:
\begin{equation}
\begin{split}
d\alpha_A = p_{Ai}\; \tilde{\omega}^i \;\;\;\; d\beta^A=q^A_i \;\tilde{\omega}^i \;\;\;\; d\omega_i = p_{Ai} \beta^A - q^A_i \alpha_A\;\;\;\; d\tilde{\omega}^i=0 \label{eq:generalized half-flat},
\end{split}
\end{equation}
where $p_{Ai}$ and $q^A_i$ are flux parameters, together with the consistency condition
\begin{equation} \label{constraint1}
p_{Ai} q^A_j - q^A_i p_{Aj} =0,
\end{equation}
coming from the identity $d^2 \omega_i=0$.

We can add for both the half-flat and the generalized cases background $H$ fluxes. These fluxes must of course satisfy the anomaly cancellation condition
\begin{align}
d H = \frac{\alpha'}{4} \left(\mathrm{Tr}(F\wedge F) - \mathrm{Tr} (R\wedge R)\right),
\end{align}
where $R$ is a suitably chosen curvature form. We shall assume that we can always choose a standard embedding so that the right hand of this equation vanishes \cite{Ali}. If so, then we can add to $H$ a background flux of the form:
\begin{align}
H_{\mathrm{fluxes}} =\mu^A \alpha_A - \epsilon_A \beta^A.
\end{align}
Then the closure of $H$ in the presence of standard embedding requires us to impose a further constraint on the flux parameters:
\be \label{constraint2}
\mu^A p_{Ai}-\epsilon_A q^A_i=0.
\ee

\subsection{The Superpotential}\label{sec:superpotential}
The superpotential $W$ at zeroth order in $\a'$ was given in eq.(\ref{eq:zero-order-superpotential}). The superpotential, to first order in $\a'$ has been computed in \cite{Gurrieri2}, and it is given by
\begin{align}
W&= (\epsilon_A - p_{Ai} T^{i}) Z^A +  (\mu^A - q^A_i T^i) \mathscr{G}_{A} + 2\a' \left(p_{ia} - q^A_i \mathscr{G}_{Aa}\right) C^i_P D^{aP}\nonumber \\
&-\frac{\a'}{3}\left[d_{ijk} j_{\bar{P}\bar{R}\bar{S}} C^{i\bar{P}}C^{j\bar{R}}C^{k\bar{S}}+\tilde{d}_{abc} j_{PRS} D^{a P}C^{b R}C^{c S}\right]. \label{eq:master-superpotential}
\end{align}
Here we give the superpotential for an generic generalized half-flat manifold. In this expression $\mathscr{G}$ represents the prepotential which in the large complex structure simply becomes,
\begin{align}
\mathscr{G}=-\frac{1}{6} \frac{\tilde{d}_{abc} Z^a Z^b Z^c}{Z^0}.
\end{align}
The constants $j_{PRS}$  and $j_{\bar{P}\bar{R}\bar{S}}$ in (\ref{eq:master-superpotential}) are the singlet pieces in the $\boldsymbol{27}\times\boldsymbol{27}\times\boldsymbol{27}$  and $\boldsymbol{\overline{27}}\times\boldsymbol{\overline{27}}\times\boldsymbol{\overline{27}}$ of $\mr{E}_6$. The fields $C^{ i \bar{P}}$ and $D^{a P}$ are the matter fields associated with the $\mathrm{E}_6$ which play a passive role in our calculations. $d_{jim}$ are the intersection numbers of the Calabi-Yau which has been deformed to make the generalized half-flat manifold, while $\tilde{d}_{abc}$ are intersection numbers on the mirror dual of the Calabi-Yau.

In our case, where we assume one dimensional complex structure moduli space, we have the complex structure index $A=0,1$ and we set $Z^0=1$ as the $Z^A$ fields are projective coordinates on the moduli space. In this case, in the large complex structure limit, the prepotential simply becomes,
\begin{align}
\mathscr{G}= -\frac{1}{6}\frac{\tilde{d}_{111} (Z^1)^3}{Z^0}.
\end{align}
Note that in our case we simply have $\tilde{d}_{111}=1$. The various derivatives of $\mathscr{G}$ that enter into (\ref{eq:master-superpotential}) can be computed from this expression keeping in mind that one sets $Z^0=1$ at the end of the calculation.
\subsection{The K\"ahler Potential for STZ}\label{sec:kahler}
In a supersymmetric theory the kinetic terms for a set of complex scalar fields $\{\Phi^i, \bar{\Phi}^{\bar{i}}\}$ are derived from the K\"ahler potential $K(\Phi^i, \bar{\Phi}^{\bar{i}})$ via
\begin{align*}
\mathcal{L}_{\mathrm{kin}}= K_{i\bar{j}} \;\partial_\mu \Phi^i \partial^\mu \bar{\Phi}^{\bar{j}},
\end{align*}
where
\begin{align*}
K_{i\bar{j}}= \frac{\partial^2 K}{\partial \Phi^i \partial \bar{\Phi}^{\bar{j}}} .
\end{align*}\\
\bl \underline{{\bf The $0^{th}$ order K\"ahler Potential} }\\\\
The K\"ahler potential, at the lowest order in $\alpha'$, is given by \cite{Gurrieri2}
\begin{align}
K = K_{\mathrm{cs}}+K_S + K_{\mathrm{K}},
\end{align}
where
\beq
K_{\mathrm{cs}} = -\ln{ \mathcal{K} \|\Omega \|^2}, \
K_{S} =- \ln{i(\bar{S} - S)}, \
K_{\mathrm{K}} = - \ln{\mathcal{K}} .
\eeq
To evaluate these K\"ahler potentials explicitly we need the explicit values of  $\|\Omega \|^2$ and $\mathcal{K}$. $\|\Omega\|^2$ is defined to be
\begin{align}
\|\Omega\|^2 =\frac{i}{\mathcal{K}} \int \Omega \wedge \bar{\Omega}. \label{eq:norm}
\end{align}
As was shown in \cite{Candelas2}  $\Omega$ is given by
\begin{align}
\Omega = Z^A \alpha_A - \mathscr{G}_A \beta^A,
\end{align}
where $\mathscr{G}_A = \frac{\partial \mathscr{G}}{\partial Z_A}$.
Using the symplectic basis to do the integral we get
\be
\int \Omega\wedge\bar{\Omega} = -Z^A \bar{\mathscr{G}}_A + \mathscr{G}_A \bar{Z}^A.
\ee
Since
\begin{align}
\mathscr{G}_0 &= \frac{1}{6} \frac{(Z^1)^3}{(Z^0)^2}=\left. \frac{1}{6} (Z^1)^3\right|_{Z^0=1}\\
\mathscr{G}_1 &= -\frac{1}{2} \left. (Z^1)^2\right|_{Z^0=1},
\end{align}
this leads to
\begin{align}
\int \Omega \wedge \bar{\Omega} &= \frac{1}{6} (Z-\bar{Z})^3, \label{eq:omegasquared}
\end{align}
where we have used the notation $Z^1\equiv Z$.\\

To evaluate (\ref{eq:norm}) we also need the following elements from the K\"ahler moduli space. $\mathcal{K}$ in the above formula is defined by:
\begin{equation}
\mathcal{K} = \frac{1}{6} \int J\wedge J\wedge J.
\end{equation}
Given that our K\"ahler moduli space is one dimensional and so
\begin{align}
J=v \omega
\end{align}
leads to
\begin{align}
\mathcal{K} =\frac{v^3}{6}.
\end{align}
Plugging this and (\ref{eq:omegasquared}) into (\ref{eq:norm}) we get
\begin{align}
\|\Omega\|^2 = \frac{i(Z-\bar{Z})^3}{v^3}.
\end{align}
These expressions given here are in terms of field variables used in reference \cite{Gurrieri2}. Their relationship to the standard variables $S$ and $T$ are given by
\begin{align*}
S  = a + i e^{-2\phi}, \ T = b+iv.
\end{align*}
Then we can reexpress the $0$-th order K\"ahler potential in terms of $S, T$ and $Z$:
\beq
K_{\mathrm{cs}} = - 3 \ln \{i (\bar{Z}-Z)\}, \
K_S = - \ln \{i (\bar{S}-S)\}, \
K_{\mathrm{K}} = - 3 \ln \{i(T-\bar{T})\}.
\eeq
\bl \underline{{\bf The $1^{st}$ order K\"ahler Potential}}\\\\
According to \cite{Gurrieri2} first order in $\alpha'$ correction to the K\"ahler potential is given by
\begin{align*}
K_{\alpha'} &= \alpha' \left[ 4 \|\Omega\|^{-2/3} g^{\mathrm{K}}_{1\bar{1}} C^{\bar{P}}C_{\bar{P}} + \|\Omega\|^{2/3} g^{\mathrm{cs}}_{1\bar{1}} D^{1 P} \bar{D}^{\bar{1}}_{\bar{P}}
-2 (K^{\mathrm{K}}_1 K^{\mathrm{cs}}_1 C^1_P D^{1 P}+ \mathrm{complex\;conjugate})\right],
\end{align*}
where $g^K_{11}$ and $g^{\mr{cs}}_{11}$ are the one dimensional metric on the K\"ahler and complex structure moduli spaces respectively.
We can then use
\beq
g^{\mathrm{K}}_{11} &=& \partial_T \partial_{\bar{T}} K_{\mathrm{K}} = \frac{-3}{(T-\bar{T})^2}, \  g^{\mathrm{cs}}_{11} = \partial_Z \partial_{\bar{Z}} K_{\mathrm{cs}} = \frac{-3}{(Z-\bar{Z})^2}\cr
K^{\mathrm{K}}_1 &=& \partial_T K_{\mathrm{K}} = \frac{-3}{(T-\bar{T})}, \
K^{\mathrm{cs}}_1 = \partial_Z K_{\mathrm{cs}}= \frac{-3}{(Z-\bar{Z})} .
\eeq
Putting all of this together we can now compute the 1st order K\"ahler potential to be:
\begin{align}
K_{\alpha'}&= -3\alpha'\left[\frac{C\bar{C}}{(Z-\bar{Z})^2} + \frac{4D\bar{D}}{(T-\bar{T})^2}+ \frac{6 (CD+ \bar{C}\bar{D})}{(T-\bar{T})(Z-\bar{Z})}\right].
\end{align}
Then the metric on the scalar manifold becomes to first order in $\a'$:
\begin{align}
K_{i\bar{j}}= \left(
\begin{array}{ccc}
 -\frac{1}{(S-\bar{S})^2} & 0 & 0 \\
 0 & -\frac{3}{(T-\bar{T})^2}  -\frac{36 \left(-2 D \bar{D}-\frac{(C D+\bar{C} \bar{D}) (T-\bar{T})}{Z-\bar{Z}}\right) \alpha' }{(T-\bar{T})^4} & \frac{18 (C D+\bar{C} \bar{D}) \alpha' }{(T-\bar{T})^2 (Z-\bar{Z})^2} \\
 0 & \frac{18 (C D+\bar{C} \bar{D}) \alpha' }{(T-\bar{T})^2 (Z-\bar{Z})^2} & -\frac{3}{(Z-\bar{Z})^2} -\frac{18 \left(-C \bar{C}-\frac{2 (C d+\bar{C} \bar{D}) (Z-\bar{Z})}{T-\bar{T}}\right) \alpha' }{(Z-\bar{Z})^4}\\
\end{array}
\right).
\end{align}
Strictly speaking we should diagonalize this matrix and then reexpress the superpotential given in the main text in terms of the redefined fields. However, here we are interested in slow moving field and thus we shall assume that the contribution from the kinetic terms are negligible, and so we can ignore the $\alpha'$ correction in the kinetic terms. Thus for the purpose of this paper we can neglect the $\alpha'$ correction to the kinetic terms and use
\begin{align}
K_{i\bar{j}}= \left(
\begin{array}{ccc}
 -\frac{1}{(S-\bar{S})^2} & 0 & 0 \\
 0 & -\frac{3}{(T-\bar{T})^2}  & 0 \\
 0 & 0 & -\frac{3}{(Z-\bar{Z})^2}
\end{array}
\right) + \mathcal{O}(\alpha').
\end{align}

\bibliography{refs}

\bibliographystyle{utphysmodb}

\end{document}